# Investigation of Magnetic Anisotropy and Heat Dissipation in Thin Films of Compensated Antiferromagnet CuMnAs by Pump-probe Experiment

M. Surýnek[1], V. Saidl[1], Z. Kašpar[1,2], V. Novák[2], R.P. Campion[3], P. Wadley[3], and P. Němec[1,*]

[1]*Faculty of Mathematics and Physics, Charles University, Ke Karlovu 3, 121 16 Prague 2, Czech Republic*
[2]*Institute of Physics ASCR, v.v.i., Cukrovarnická 10, 162 53 Prague 6, Czech Republic*
[3]*School of Physics and Astronomy, University of Nottingham, Nottingham NG7 2RD, United Kingdom*

We recently reported on a method to determine the easy axis position in a 10 nm thick film of the fully compensated antiferromagnet CuMnAs. The film had a uniaxial magnetic anisotropy and the technique utilized a magneto-optical pump and probe experiment [Nature Photonics 11, 91 (2017)]. In this contribution we discuss the applicability of this method for the investigation of a broader set of epitaxial CuMnAs films having different thicknesses. This work reveals that the equilibrium magnetic anisotropy can be studied only in samples where this anisotropy is rather strong. However, in the majority of CuMnAs films, the impact of a strong pump pulse induces nano-fragmentation of the magnetic domains and, therefore, the magnetic anisotropy measured by the pump-probe technique differs substantially from that in the equilibrium conditions. We also demonstrate that optical pump-probe experiment can be used very efficiently to study the local heating and heat dissipation in CuMnAs epitaxial layers. In particular, we determined the electron-phonon relaxation time in CuMnAs. We also observed that for a local film heating by a focused laser the thinner films are heated more, but the heat is dissipated considerably faster than in the case of thicker films. This illustrates that the optical pump-probe experiment is a valuable characterization tool for the heat management optimization in the CuMnAs memory devices and can be applied in a similar way to those used during heat-assisted magnetic recording (HAMR) technology development for the latest generation of hard drive disks.

## I. INTRODUCTION

Antiferromagnets (AFs) are very promising materials for spintronic applications thanks to the many interesting properties they combine [1-4]. For instance, the absence of net magnetization and stray fields eliminates crosstalk between neighboring devices that enables their high-density arrangement and makes them robust against external magnetic fields. Moreover, due to the presence of strong exchange coupling between the sublattices in AFs, the intrinsic resonance frequencies are in the terahertz frequency range, in contrast to gigahertz frequencies in ferromagnets (FMs), which offers the prospect of extremely fast operation of the devices [5]. On the other hand, the absence of a net magnetic moment, the frequently observed small size of magnetic domains and the ultrafast magnetization dynamics make probing of antiferromagnetic order by common magnetometers or magnetic resonance techniques notoriously difficult. Therefore, the possibility to study AFs by optical techniques attracts significant attention nowadays [6]. Moreover, optics can be used not only to study AFs passively, but also to actively manipulate its magnetic order [7,8]. Recent breakthroughs in electrical detection and manipulation of antiferromagnets have opened a new avenue in the research of non-volatile spintronic devices. The first experimental realization of all-electrical room-temperature USB-compatible antiferromagnetic memory chips was reported in antiferromagnetic metal CuMnAs [9]. It was also demonstrated that writing in this CuMnAs-based memory device can be achieved by free propagating THz pulses, which have a picosecond duration [10]. We recently developed an experimental technique allowing us to determine the easy axis position in 10 nm thick film of compensated antiferromagnet CuMnAs, possesing in-plane uniaxial magnetic anisotropy, by magneto-optical pump and probe experiment [11]. In this contribution we discuss the applicability of this method for the investigation of a broader set of epitaxial MBE-grown CuMnAs samples having different thicknesses. Moreover, we demonstrate that optical pump-probe signals can be very efficiently used to study the local heating and heat dissipation in CuMnAs films.

[*] Electronic mail: nemec@karlov.mff.cuni.cz

## II. EXPERIMENTAL TECHNIQUE AND SAMPLES

We employed the pump-probe technique of ultrafast laser spectroscopy. As the light source was used a tunable femtosecond Ti:Sapphire oscillator (Mai Tai, Spectra Physics) that produces laser pulses with time-width of 150 fs and has a repetition rate of 80 MHz. In the standard (degenerate) pump-probe experiments, the laser output was divided into strong pump pulses with a fluence of $\approx$ 3 mJ cm$^{-2}$ and probe pulses with at least 50 times weaker fluence. For the non-degenerate pump-probe experiment, when pump and probe pulses have different wavelengths, the optical parametric oscillator (Inspire, Spectra Physics) was used as a source of probe pulses and depleted fundamental beam was used as the pump beam. In all experiments we used laser pulses with photon energies smaller than the band gap energy of GaP substrate (see also Fig. 6). The time delay between pump and probe pulses was set by a computer-controlled delay line in the pump beam (see also Fig. 8). Typically, experiments were performed in nearly normal incidence geometry with an angle between pump and probe beam of 6° when pump and probe pulses were focused to a same spot on the sample with a spot-size of ≈ 30 μm (full width at half maximum, FWHM). Control experiments were also performed with collinearly propagating pump and probe pulses that were focused by a microscopic objective to a laser spot-size of ≈ 1.5 μm [12,13]. The samples were mounted in a closed-cycle helium cryostat (ARS) and the experiments were performed at a base temperature of 15 K. Further experimental details about the used setup and method are provided in Appendix B. In transmission and reflection geometries we measured simultaneously the magnetooptical (MO) signal, corresponding to the probe polarization rotation, and change of probe intensity, corresponding to differential reflectivity $dR/R$ and differential transmission $dT/T$ for reflected and transmitted probe pulses, respectively. If the changes in transmission are small, the differential transmission $dT/T \sim \Delta\alpha d$, where $d$ is the sample thickness and $\Delta\alpha$ is the pump-induced change of the absorption coefficient [14].

In this study we investigated the fully compensated collinear antiferromagnetic semimetal CuMnAs. A broad set of CuMnAs thin epitaxial films, with layer thicknesses varying from 8 to 60 nm, were grown on GaP(100) substrate by molecular beam epitaxy both in Prague (Institute of Physics, Academy of Sciences of the Czech Republic) and Nottingham (School of Physics and Astronomy, University of Nottingham). To prevent the oxidation of CuMnAs surface, a protective ≈ 2 nm Al cap was grown on each CuMnAs epilayer. The magnetic moments lie in the basal plane of the CuMnAs films and its tetragonal lattice is matched to GaP(100) by 45° rotation (CuMnAs{100} is parallel to GaP{110}) [15]. The Néel magnetic ordering temperature in CuMnAs films is around 480 K [11, 16]. Previous detailed studies of magnetic anisotropy of CuMnAs films by X-ray magnetic linear dichroism (XMLD) spectroscopy revealed that films with a thickness of about 10 nm have uniaxial magnetic anisotropy along the [110] crystallographic direction of the GaP substrate [11,16], which corresponds to the [100] direction in CuMnAs. In contrast, thicker films show biaxial magnetic anisotropy along the [100] and [010] directions in CuMnAs [17,18]. Optical properties of the studied samples are described in detail in Appendix A.

## III. RESULTS AND DISCUSSION
### A. TIME RESOLVED MAGNETO-OPTICAL EXPERIMENT

In the fully compensated collinear antiferromagnet CuMnAs, the signals from oppositely oriented magnetic sublattices cancel in equilibrium for MO effects which are linear in magnetization (i.e., Kerr effect and Faraday effect). Hence, the Voigt effect in absorbing medium, which is a MO effect that is quadratic (even) in magnetization, is a very useful tool to study this material [6]. This effect leads to a rotation of the polarization plane of linearly polarized light (or change of its ellipticity) when the magnetic moments are perpendicular to the light propagation direction. Phenomenologically, it can be described by a different complex index of refraction for light with a polarization plane oriented parallel and perpendicular to the magnetic moments, respectively [19,20]. The polarization rotation due to Voigt effect in absorbing medium is typically rather small in comparison with other (non-magnetic) sources of light polarization changes [8,21,22]. In order to separate the actual MO signal from non-magnetic contributions to the total polarization rotation experienced by light in the investigated sample, we utilized pump-probe technique [11], as schematically depicted in Fig. 1a. The pump-induced local heating of the sample leads to its partial demagnetization [23,24] and, consequently, to a reduction of the MO signal. The measured dynamical MO signal in pump-probe experiment is given by [11]

$$MO(\Delta t, \epsilon) = \frac{2P}{M}\sin 2(\varphi - \epsilon)\,\delta M(\Delta t) + D \quad (1)$$

where $P$ corresponds to MO coefficient of Voigt effect in absorbing medium, which is scaling quadratically with the sublattice magnetization $M$ projection onto the plane perpendicular to the probe light propagation direction, and $\varphi$ and $\varepsilon$ describe the in-plane orientation of magnetic moments and light polarization, respectively. $\delta M$ is a pump-induced change of magnetization, which depends on time delay $\Delta t$ between pump and probe pulses, and $D$ is a polarization-insensitive background, which is coming mainly from the substrate.

As a typical example, we show in Fig. 1b the time-resolved dynamics of pump-induced change of MO signal in 10 nm thick epilayer of CuMnAs in transmission geometry for 3 selected orientations of incident light polarization plane $\varepsilon$. At time delay $\Delta t = 0$ ps the strong pump pulse locally heats the sample (see Fig. 5) causing its partial demagnetization. Consequently, the probe pulses measure the pump-induced polarization rotation change with a characteristic harmonic dependence on $\varepsilon$, which is described by Eq. (1). A detailed polarization dependence of MO signal at time delay $\Delta t = 60$ ps is shown in Fig. 1c.

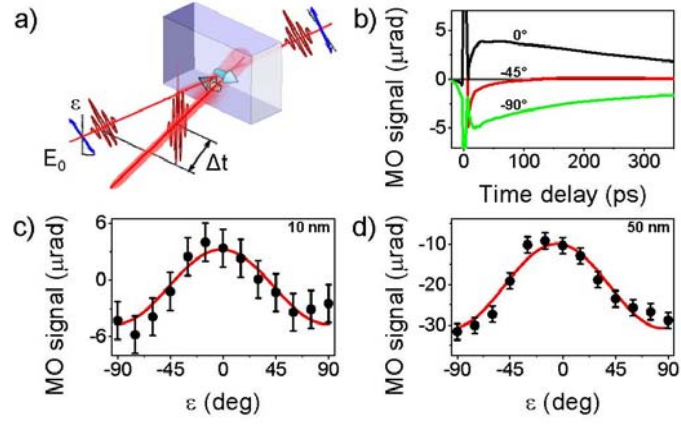

**Figure 1: Pump-probe magneto-optical experiment in CuMnAs.** (a) Schematic illustration of the pump-induced reduction of Voigt effect MO signal that is measured for various incident probe polarization orientations $\varepsilon$ as a function of time delay $\Delta t$ between pump and probe pulses. (b) Time-resolved MO data measured in transmission geometry in 10 nm thick film grown in Prague using pump and probe pulses with wavelength of 820 nm. (c) Polarization dependence of MO signal at $\Delta t = 60$ ps for the same sample as in (b); polarization orientation $\varepsilon = -45°$ corresponds to the crystallographic direction [110] in the GaP substrate. (d) Same dependence as in (c) for 50 nm thick film grown in Prague.

As expected [11], the deduced polarization dependence for this 10 nm epilayer has symmetry corresponding to uniaxial sample with an easy axis located along one of the polarization orientations where the measured MO signal is "zero" (i.e., for $\varepsilon = \pm 45°$ in this case). If needed, this ambiguity of the Néel vector orientation determination can be removed by an additional experiment where tilting of the sample around these two directions is performed (see Fig. 2 in Ref. 11). The situation for thicker films, which have biaxial magnetic anisotropy, is more complicated. If the magnetic domains are larger than the laser spot-size and the laser spot is located within a single magnetic domain, all the magnetic moments sensed by the laser would be pointing along the same direction and the detected MO signal would have again the symmetry predicted by Eq. (1). Consequently, for two orthogonal domains the MO signals would have the opposite sign. However, as the domain size of biaxial CuMnAs films does not exceed a few micrometers [17,18], which is considerably less than the laser spot-size of $\approx 30$ μm in this experiment, several domains should be sampled simultaneously by the laser beam and, consequently, the detected MO signals should average out to zero. On the other hand, in a control experiment (not shown here) with a spot-size of $\approx 1.5$ μm, which is comparable or even smaller than the biaxial domain size in CuMnAs films with certain thicknesses (see, e.g., Fig. 4d in Ref. 25), we expected to see a strongly position-dependent MO signal. In reality, however, both experiments with different spot sizes provided rather similar results that are demonstrated by Fig. 1d where MO data measured in 50 nm epilayer are shown. Overall, very similar results, both from the symmetry point of view and also with respect to the magnitude of detected MO signals, were obtained in all studied samples (with a thickness from 8 to 60 nm) and for two different spot-sizes (30 and 1.5 μm) for many different positions on the samples.

In the following, we discuss the origin of this discrepancy between the results obtained by XMLD, where a change of CuMnAs magnetic anisotropy from uniaxial [11,16] to biaxial [17,18] was observed when the epilayer thickness was increased from ≈ 10 to 50 nm, respectively, and our pump-probe experiments, where very similar results were obtained for all the investigated samples with thicknesses varying from 8 to 60 nm even in a case when the laser spot size was smaller than the magnetic domain size. XMLD is expected to detect the actual magnetic anisotropy of the CuMnAs epilayers in experimental conditions close to the thermodynamic equilibrium. In contrast, the inherent part of the pump–probe technique is a local transient heating of the sample by a pump pulse with a rather large temperature increase (see Fig. 5). The magnetic anisotropy and stability of domain structure in the volume of CuMnAs films is apparently rather weak and, therefore, this abrupt temperature increase seems to lead to nano-fragmentation of magnetic domains within the irradiated spot, as schematically depicted in Fig. 2, where the MO signal is averaged out to zero. The remaining observable MO signal, which corresponds to uniaxial magnetic anisotropy and which we observed in all the studied samples (see Figs. 1c and 1d), is probably the (relatively strong) uniaxial magnetic anisotropy induced in the interfacial layer of CuMnAs that comes from the specific symmetry of bond alignments on the CuMnAs/GaP interface [11]. Unlike in the pump-probe optical experiment, which is typically performed in transmission geometry, the XMLD experiment was performed using total electron yield (TEY), a surface sensitive technique relying on the emission of Auger electrons, that leads to a probing depth of ≈ 3-5 nm, which is smaller than the typical CuMnAs film thickness. Consequently, this uniaxial region close to the CuMnAs/GaP interface is generally not visible to TEY-XMLD. This explanation is also fully in accord with our previous observation that in the Fe/CuMnAs bilayer we did not observe in the pump-probe experiment, unlike in XMLD experiment, any significant reorientation of magnetic moments in CuMnAs by external magnetic field due to the interlayer exchange coupling with Fe [26]. In fact, very recently, the nano-fragmentation of magnetic domains due to the film local heating was suggested [25] as an origin of the large electrical readout signals measured in CuMnAs memory devices after strong electrical and optical writing pulse. This nano-fragmentation of domains was also directly imaged by X-ray magnetic linear dichroism - photoemission electron microscopy (XMLD-PEEM) (see Fig. 4d in Ref. 25), magnetometry using NV centers in diamond (see Figs. 3 and 4 in Ref. 27) and magneto-Seebeck microscopy (see Fig. 4 in Ref. 28).

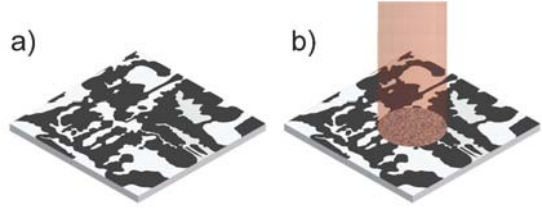

**Figure 2: Schematic illustration of the magnetic domain structure change due to pump-induced heating of the sample.** (a) Assumed fictional equilibrium domain structure of the CuMnAs films, which resembles the real domain structures observed by XMLD-PEEM in thicker films with a biaxial magnetic anisotropy [17,18]. (b) Assumed local change of the domain structure due to the laser.

### B. TRANSIENT REFLECTIVITY AND ABSORPTION MEASUREMENT

In the previous section we have shown that pump-probe *magneto-optical* experiments do not provide much information about the magnetic anisotropy in CuMnAs films in equilibrium. In this section we demonstrate that pump-probe *optical* experiments can be used very efficiently to study dynamics of heating and heat dissipation in these films.

The photoexcitation of a metal by an intense femtosecond laser pulse excites the electron distribution out of equilibrium on a time scale much shorter than the electron-phonon interaction time. The resulting non-thermal population of electrons thermalizes rapidly by electron-electron scattering processes. Consequently, a thermalized electron system, which can be described by a Fermi distribution with an electron temperature $T_e$, is formed within ≈ 100 fs after the impact of the pump pulse [29-31]. On a picosecond timescale, the excess energy is dissipated from the electron system to the lattice by electron-phonon scattering processes, which leads to an increase of the lattice temperature $T_{lat}$. Due to the energy transfer from the electrons to the phonons (and magnons) a thermal stress is generated that can launch coherent acoustic phonon wavepackets (strain waves). These acoustic wavepackets propagate into the substrate with an acoustic sound velocity $v_s$ (in GaP $v_s$ ≈ 6 km s$^{-1}$; see also Eq. (S16) and Fig. S10 in the Supplementary information of Ref. 11) and their presence leads to oscillations in the measured transient reflectivity signals – see Fig. 3a. Finally, on a longer time scale, the heat diffusion dissipates the excess energy and the metal returns to the equilibrium state. Importantly, all the above effects lead to a change of optical properties and, therefore, the corresponding characteristic time constants can be evaluated from the measured optical transient signals [29-31]. Experimentally, this is usually achieved by a degenerate pump-probe experiment where a time evolution of differential reflectivity d$R/R$ is measured by probe pulses of the same wavelength ($\lambda$) as that of the pump pulse [30]. The measured reflectivity dynamics can be reproduced by a phenomenological equation [31]

$$\Delta R/R (\Delta t, \lambda) = \left[\alpha(\lambda)\left(1 - e^{-\Delta t/\tau_{ee}}\right)e^{-\Delta t/\tau_{ep}} + \beta(\lambda)\left(1 - e^{-\Delta t/\tau_{ep}}\right)\right]e^{-\Delta t/\tau_{th}}. \quad (2)$$

The first term, with a spectral weight $\alpha$, represents the electronic response with a rise time described by the electron-electron thermalization time $\tau_{ee}$ and decaying by an energy transfer to the lattice with the characteristic electron-phonon relaxation time $\tau_{ep}$. The second term, with a spectral weight $\beta$, describes the lattice heating, with the same time constant time $\tau_{ep}$, and the thermal relaxation time $\tau_{th}$ represents the heat diffusion from the excited area. The heat capacity of the electron system is much smaller than the heat capacity of the lattice. Consequently, for $\Delta t \gg \tau_{ep}$, when $T_e = T_{lat}$, almost all the excess energy is stored in the lattice system and the contribution from the electron system to the measured change of optical properties is negligible [29-31]. In general, a detailed interpretation of the differential reflectivity can be very complex because the magnitude and even sign of $\alpha$ and $\beta$ depend strongly on details of the electronic band structure and wavelength of probe pulses [29-31] (see also Fig. 3). Consequently, for an unambiguous separation of individual contributions to the measured signals, it is very advantageous to measure the differential reflectivity at several probe wavelengths [29,31] and/or to supplement it with measurements of the differential transmission d$T/T$ [29].

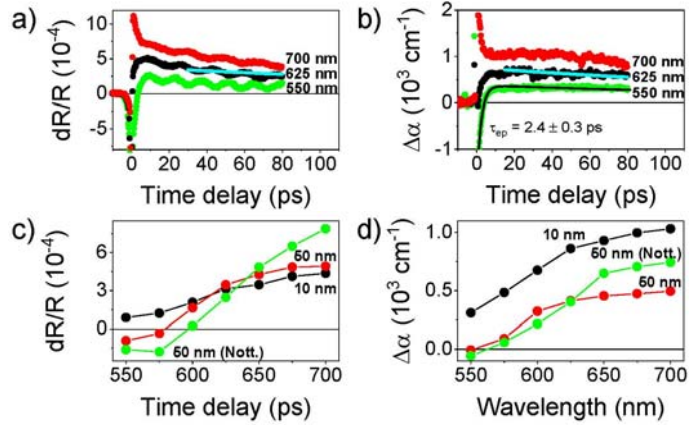

**Figure 3:** (a) and (b) Time-resolved dynamics of differential reflectivity d$R/R$ (a) and absorption coefficient change $\Delta\alpha$ (b) measured in 10 nm thick film grown in Prague for 3 selected wavelengths of probe pulses after excitation by pump pulses at 820 nm (points). Cian lines are monoexponential decay fits providing the thermal relaxation time $\tau$th (see Fig. 4c); black line in (b) is a fit by Eq. (2) yielding also the electron-phonon relaxation time $\tau$ep = 2.4 ± 0.3 ps. (c) and (d) Spectral dependence of d$R/R$ (c) and $\Delta\alpha$ (d) at $\Delta t$ = 60 ps (i.e., signals coming from the lattice) for 10 nm and 50 nm films grown in Prague and for 50 nm film grown in Nottingham.

In Fig. 3a and Fig. 3b we show the dynamics of differential reflectivity d$R$/$R$ and pump-induced change of absorption coefficient $\Delta\alpha$, respectively, that was measured in a non-degenerate pump-probe experiment by probe pulses of several wavelengths in 10 nm film prepared in Prague. The strong wavelength dependence of the initial dynamics (the sign change, in particular) clearly identifies the contribution coming from the electron system. Consequently, it shows that the temperature of electrons and lattice is equilibrated after $\Delta t \approx$ 10 ps (where this strongly wavelength dependent signal from the electron system is no longer present). Rather similar results were obtained also for films with other thicknesses. These experiments enabled us to determine the electron-phonon relaxation time in CuMnAs $\tau_{ep}$ = 2.4 ± 0.3 ps, which can be compared with the characteristic times reported for Au ($\tau_{ep}$ = 1 ps [29]), Ni (0.4 ps [30]) and Fe (1.4 ps [30]). The measured spectral dependences of transient optical signals coming from the lattice (for $\Delta t$ = 60 ps), which are due to a complex index of refraction change induced by the lattice temperature increase, are shown in Fig. 3c and Fig. 3d for two different film thicknesses. The measured data reveal the very similar optical (and also thermal, see Fig. 4b and 4c) properties of CuMnAs films grown in Prague and Nottingham and, therefore, points to a high reproducibility of the film growth process by MBE. However, the most important conclusion within the context of this paper is that it is possible to determine the dynamics of heat dissipation from the measured data, which corresponds to the monoexponential decay fits of the measured dynamics for long time delays [cf. Eq. (2)]. Interestingly, the fitting procedure is easier to perform in the measured dynamics of $\Delta\alpha$ than in d$R$/$R$ where the thermal relaxation-related signal dominates after longer time delays (see Figs. 3a and 3b). Moreover, the fitting of d$R$/$R$ is also complicated by the presence of an oscillatory signal that is due to the propagation of coherent acoustic phonon wavepackets in the GaP substrate (see Figs. S9 and S10 and the adjacent discussion in the Supplementary information of Ref. 11).

The heat dissipation dynamics in films of different thicknesses is compared in Fig. 4. The measured data clearly show that the heat is dissipated faster in thinner films – see Fig. 4c. Moreover, assuming that the change of optical constants is linear in the lattice temperature [30], our data suggest that the pump-induced local heating of the CuMnAs lattice is larger in thinner films – see Fig. 4b. To achieve a deeper understanding of these results, we performed numerical heat transfer simulations, the results of which are shown in Fig. 5 (see Appendix C for details). These simulations match nicely the experimentally measured dynamics shown in Fig. 4. that enabled us to deduce the values of the initial CuMnAs temperature increase $\Delta T$ for films with different thicknesses – see Fig. 5b. These data confirm that thinner films are heated considerably more by pump laser pulses than the thicker films, which is a consequence of their different absoptance depth profiles (see Fig. 7 in Appendix A).

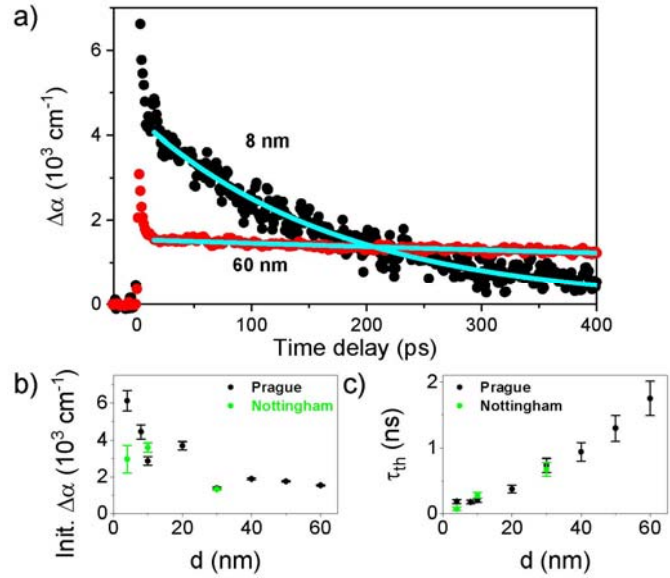

**Figure 4: Experimental measurement of local heating and heat dissipation in CuMnAs films of different thicknesses.** (a) Dynamics of $\Delta\alpha$ measured for 8 nm and 60 nm films grown in Prague by pump and probe pulses with wavelength of 820 nm (points); lines depict the monoexponential decay fits revealing the thermal relaxation time $\tau_{th}$. (b) and (c) Dependence of initial value of $\Delta\alpha$ (b) and thermal relaxation time $\tau_{th}$ (c) on epilayer thickness for samples grown in Prague and Nottingham.

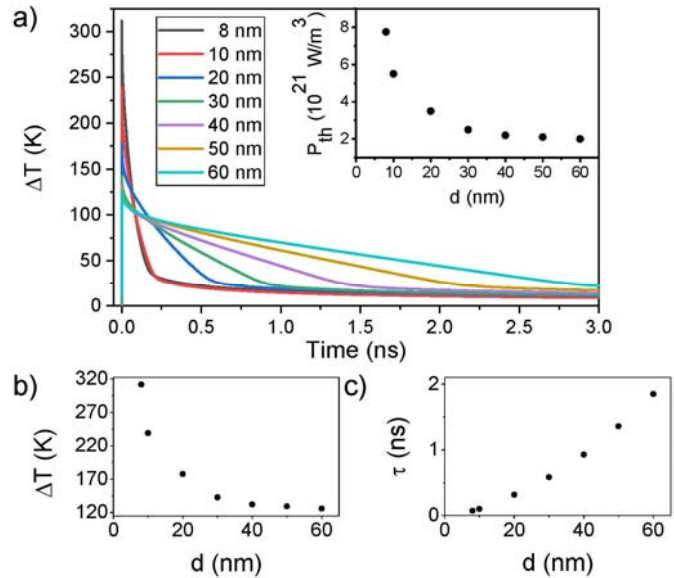

**Figure 5: Heat transfer simulations in CuMnAs films of different thicknesses.** (a) Transient change of CuMnAs temperature increase $\Delta T$, relative to the sample base temperature of 15 K, computed for denoted film thicknesses. Inset: Absorbed effective power density used in simulations for individual film thicknesses (see Appendix C). (b) and (c) Dependence of initial value of $\Delta T$ (b), and effective thermal relaxation time $\tau$ (c) on film thickness. Note that at elevated sample base temperature the values of $\Delta T$ would be smaller due to a strong temperature dependence of the heat capacity (see Fig. 9).

## IV. CONCLUSIONS

In this contribution we investigated the research potential of pump-probe experiments for the investigation of a compensated metal antiferromagnet CuMnAs. We revealed that the previously demonstrated technique for studying magnetic anisotropy, which is based on measurement of magneto-optical pump-probe signals (Ref. 11), is applicable only for samples where the uniaxial magnetic anisotropy is rather strong. In samples with a weak magnetic anisotropy, which seems to be the case for a vast majority of CuMnAs epitaxial layers, the impact of a strong pump pulse itself induces considerable changes of the magnetic anisotropy and, therefore, the magnetic anisotropy measured by the pump-probe technique differs substantially from that in the equilibrium conditions. Nevertheless, the assumed pump-induced nano-fragmentation of magnetic domains seems to be closely connected with the recently observed large electrical readout signals measured in CuMnAs memory devices after strong electrical and optical excitation (Refs. 25,27,28). We also demonstrated that optical pump-probe signals can be very efficiently used to study the heating and heat dissipation in CuMnAs films. In particular, we determined the electron-phonon relaxation time in CuMnAs ($\tau_{ep} = 2.4 \pm 0.3$ ps). Moreover, we observed that thinner CuMnAs films are heated considerably more by femtosecond laser pulses than the thicker films but the excess heat is dissipated considerably faster from them. This illustrates that optical pump-probe experiment could be used very efficiently as a characterization tool for the heat management optimization in the CuMnAs memory devices in a similar way that it was applied during the heat-assisted magnetic recording (HAMR) technology development for the recent generation of the hard drive disks (HDDs) [30]. For example, the pump-probe studies in HDDs enabled to achieve an extremely fast cooling of the recorded bits due to an efficient vertical heat diffusion using an optimized heat sink underlayer structure (see Fig. 18 in Ref. 32). Consequently, detailed spatially- and time-resolved pump-probe studies of heat dissipation in CuMnAs films on various substrates can be extremely important for the future development of memory devices from this material.

## ACKNOWLEDGMENTS

This work was supported in part by the Grant Agency of the Czech Republic under EXPRO grant no. 19-28375X, by EU FET Open RIA under grant no. 766566, from the Charles University grants no. 1582417 and no. SVV-2019-260445, the Engineering and Physical Sciences Research Council under grant no. EP/P019749/1, and by the Ministry of Education of the Czech Republic Grant LM2018110. P.W. acknowledges support from the Royal Society through a University Research Fellowship.

**AVAILABILITY OF DATA**

The data that support the findings of this study are available from the corresponding author upon reasonable request.

**APPENDIX A: OPTICAL PROPERTIES OF STUDIED CUMNAS FILMS**

The studied samples were prepared by molecular beam epitaxy of antiferromagnetic metal CuMnAs on 0.5 mm thick GaP substrate, which is a semiconductor with a band gap of about 2.3 eV [33], with a 2 nm thick Al capping layer [15]. The capping layer is naturally oxidized when the sample is exposed to air forming a protective layer that is fully transparent in the relevant spectral range and, therefore, it will be neglected in the following discussion for simplicity.

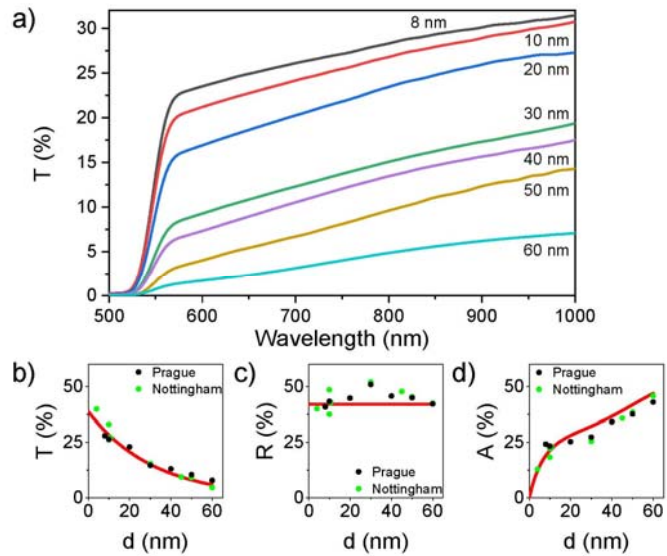

**Figure 6: Optical characterization of CuMnAs films with different thicknesses.** (a) Spectral dependence of films transmittance T; the change of T below 550 nm is due to absorption of GaP substrate. (b) and (c) Experimentally measured values of transmittance $T$ and reflectance $R$, from the air/CuMnAs film side, at ≈ 800 nm (points); lines depict the expected dependence on the film thickness (see text). (d) Thickness dependence of film absorptance $A$, which was deduced from the measured values of $T$ and $R$ (points). Lines represent the computed values of $A$ that were obtained by integrating the depth profiles shown in Fig. 7.

In Fig. 6a we show the spectra of transmittance that were measured in samples containing CuMnAs epilayers with different thicknesses $d$ (light was incident on the samples from the air/CuMnAs side). For wavelengths longer than ≈ 550 nm, where the GaP substrate is transparent, a systematic dependence of the sample transmittance on the CuMnAs film thickness is clearly apparent. The sample transmittance $T$, which is defined as a ratio between the transmitted ($I_T$) and incident ($I_0$) light intensities, is affected by the absorption of light in the CuMnAs film and by the sample reflectance $R$, which is defined as a ratio between the reflected ($I_R$) and incident ($I_0$) light intensities. To disentangle these effects, we measured (close to a normal angle of incidence) $I_T$ and $I_R$ for samples with different $d$ using light with wavelength of ≈ 800 nm, from which the dependence of $T$ and $R$ on $d$ was evaluated – see Figs. 6b and 6c, respectively. In the studied samples, there are three interfaces that can reflect light: air/CuMnAs, CuMnAs/GaP and GaP/air. The corresponding reflectances, which were computed from the indices of refraction at 800 nm ($n^{CuMnAs}$ = 3.5 + i 2.0 [34] and $n^{GaP}$ = 3.2 [33]), are $R_1$ = 42%, $R_2$ = 8.4% and $R_3$ = 27%, respectively. The experimentally measured reflectances, which are shown in Fig. 6c, do not depend systematically on $d$ and agree well with the value of reflectivity $R_1$ of the air/CuMnAs surface, as expected. Similarly, the measured transmittance agrees well with the expected dependence expressed by Eq. (A1)

$$T = (1 - R_1)(1 - R_2)(1 - R_3)e^{-\alpha d} , \tag{A1}$$

where the absorption coefficient of the CuMnAs $\alpha = 3.1 \times 10^5$ cm$^{-1}$ was computed from the imaginary part of $n^{CuMnAs}$. Finally, the absorptance $A$, i.e., the fraction of light intensity that is absorbed in the CuMnAs layer, can be evaluated as $A = 1 - T - R - S$, where $S$ describes the light scattering at imperfect surfaces. Using $S \approx 7\%$, which we estimated from the reference measurement using a bare GaP substrate, which is assumed to be non-absorbing at 800 nm, we obtained the required dependence of $A$ on the CuMnAs thickness – see Fig. 6d. Surprisingly at first sight, the amount of absorbed light *does not* depend exponentially on the film thickness, as might be expected from the Beer-Lambert law. The reason is that also (counter-propagating) light that was reflected at the CuMnAs/GaP and GaP/air interfaces is absorbed in thinner CuMnAs films. (We note that this is the origin of the second peak that is observed in time-resolved signals in these samples for time delay $\Delta t \approx 10$ ps – see Fig. S10e in the Supplementary information of Ref. 11.) To include these effects in calculations, we used a transfer-matrix method optics package *tmm* for Python software [35]. The obtained absorptance depth profiles are shown in Fig. 7 and the total absorptances, which were obtained by integrating these profiles, are depicted as solid line in Fig. 6d, in an excellent agreement with the measured values of $A$.

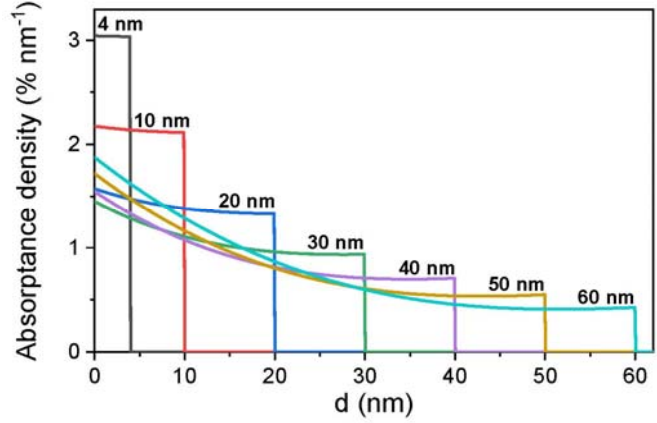

**Figure 7: Computed depth profiles of absorptance density for different thicknesses of CuMnAs films on GaP substrate.** To remove the intereference effects from the computed profiles, we averaged the ensemble of 1000 coherent simulations [35] with varying thickness of GaP substrate (from 499.6 to 500.4 μm).

**APPENDIX B: EXPERIMENTAL DETAILS ABOUT PUMP-PROBE EXPERIMENT**

A schematic depiction of the pump-probe setup used in reflection geometry is shown in Fig. 8 [36]. Femtosecond laser pulses are split by a beam splitter (*BS*) to pump and probe pulses with a delay line (*DL*) controlled mutual time delay ($\Delta t$). The intensity of the pump beam is modulated at a frequency $f \approx 2$ kHz by an optical chopper (*CH*). The required orientation of polarization plane of linearly-polarized laser pulses is set using a combination of polarizers (*P*) and half wave plates ($\lambda/2$). The pulses are focused by a single lens (*L*) on the sample (*S*) in a cryostat, which is placed between the poles of an electromagnet (*EM*) producing external magnetic field ***B***. The reflected probe pulses are collimated by another lens and directed toward an optical bridge. Here the probe beam is split by a Glan-Laser (*GL*) polarizer to two arms having *s* and *p* linear polarizations and the corresponding light intensities $I_s$ and $I_p$ are measured by two silicon photodiode detectors. The corresponding electrical signals are then added (for the reflectivity measurement) and subtracted (for the MO signal measurement), amplified and processed by lock-in amplifiers, which extract the signal modulated at the chopper-frequency. Before the actual collection of time-resolved data, the bridge has to be balanced with the pump beam blocked ("*off*" state in the following). This balancing is accomplished by moving the chopper to the probe beam and rotating the half wave plate in front of the *GL* until the "difference signal" ($I_s^{off} - I_p^{off}$) is zero, which represents a reference equilibrium state without the sample excitation. At this stage, the corresponding equilibrium "sum signal" ($I_s^{off} + I_p^{off}$) is also recorded, which is necessary for the evaluation of the measured signals magnitude (see Eqs. (A2) and (A3) below). Next, the chopper is placed back to the pump beam path and (with a probe beam blocked) the lockin phase is set in such a way that scattered pump photons produce maximum positive "sum signal". Finally, the probe beam is unblocked. When the sample is excited by the pump pulse, the equilibrium sample reflectivity *R* is changed and, therefore, the dependence of the pump-induced change of the "sum signal" ($\Delta I_s + \Delta I_p$) can measured as a function of a time delay $\Delta t$. Consequently, the time evolution of the differential reflectivity d*R*/*R* is given by [37]

$$\frac{dR}{R}(\Delta t) = \frac{(\Delta I_s + \Delta I_p)(\Delta t)}{I_s^{off} + I_p^{off}} \ . \tag{A2}$$

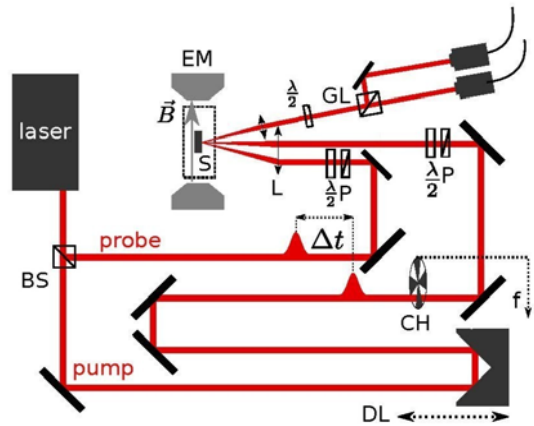

**Figure 8: A schematic depiction of the used pump-probe setup used in reflection geometry.**

Simultaneously, the impact of pump pulse can change the magnetic properties of the sample which leads (via MO effects) to a change of probe polarization. Consequently, this shifts the bridge off the balance and the pump-induced change of the "difference signal" ($\Delta I_s - \Delta I_p$) can be measured as a function of $\Delta t$. If the measured polarization rotation and reflectivity change are small, the pump-induced polarization rotation $\Delta\theta$ is given by [37]

$$\Delta\theta(\Delta t) \approx \frac{(\Delta I_s - \Delta I_p)(\Delta t)}{2\left(I_s^{off} + I_p^{off}\right)} \ . \tag{A3}$$

For experiments in transmission geometry the optical setup and data processing are quite analogous only transmitted probe photons are detected and the corresponding "sum signal" corresponds to the differential transmission dT/T.

In general, it is better to use a probe beam size on the sample that is smaller than that of a pump beam otherwise the probe beam averages signals for different excitation fluences. On the other hand, similar sizes of pump and probe beams are advantageous for a good signal-to-noise ratio in degenerate pump-probe experiment, where scattered pump photons are a dominant source of a noise in the measured signals. Consequently, in our particular case, where the measured signals do not depend super-linearly on the pump fluence, we decided to use the same sizes of pump and probe beams and we confirmed experimentally that the measured dynamics was not affected significantly by the size of the beams.

The major advantage of measuring this pump-induced change of MO signal relative to the case of more common measurement of static MO signal is that it enables a separation of light polarization rotation due to the Voigt effect in absorbing medium from that of a non-magnetic origin (e.g., due to a strain in cryostat windows or a crystal structure of the GaP substrate). The key ingredient is that any polarization changes experienced by probe pulses during their propagation in the optical setup that are not modified by the pump pulses are not detected by this technique. Hence, it is sensitive only to changes occurring in the $\approx$ 100 μm region where the pump and probe beams spatially overlap (see Fig. 8). In this sample part, the pump-induced local heating of the CuMnAs epilayer leads to its partial demagnetization and, consequently, to a reduction of the static MO signal (see Fig. 4 in Ref. 11) which is subsequently measured by probe pulses. In principle, even the measured dynamical MO signals can contain information not only about the pump-induced magnetization change but also about the pump-induced change of the complex index of refraction (so-called "optical part" of the MO signal) [38]. Consequently, the dynamics of both rotation and ellipticity has to be measured and compared before the obtained MO signal can be attributed to the magnetization dynamics. As demonstrated in Ref. 11, see the supplementary Eqs. (s9) – (s12) and Fig. S4, the experimentally measured dynamics of pump-induced change of the rotation and elipticity is usually identical for time delays larger than 20 ps, which confirms the dominance of "magnetic signal" in the measured MO data in CuMnAs films. This conclusion is also corroborated by the measured temperature dependence of the measured dynamical MO data from which the Néel magnetic ordering temperature $T_N$, which agrees with that obtained from independent electrical experiments, can be obtained (see Fig. 4 in Ref. 11). Finally, as demonstrated in Ref. 11, the conclusions about uniaxial magnetic anisotropy of thin CuMnAs films that were deduced from the pump-probe MO experiment are fully in accord with that obtained by the well-established XMLD technique. Nevertheless, for certain samples and/or experimental conditions there might be also a polarization rotation due to pump-induced change of optical anisotropy (i.e., changes of the optical indicatrix in the CuMnAs film and/or the GaP substrate). However, these signals can be usually separated by their independence on the polarization of probe pulses [term $D$ in Eq. (1)] and/or the sample temperature (see Fig. 4 in Ref. 11).

Finally, we would like to mention why we are not using in our analysis of the measured dynamical data the well-established three-temperature-model [23,24] which is frequently used to describe simultaneously the magnetic, electron and lattice dynamics in pump-probe experiments in magnetic materials. Our research was motivated by the memory applications of CuMnAs thin films and, therefore, our aim was to study the properties of these films that are directly connected with the resistivity changes, which are induced in this material system by electrical writing pulses. From this perspective, the magnetic anisotropy and heat dynamics in CuMnAs films where the properties that were the most relevant for a construction of CuMnAs-based memory devices. On the other hand, the demagnetization dynamics does not seem to be of that importance for the electric functionality optimization of the memory device performance and, therefore, we did not study these effects in detail.

**APPENDIX C: HEAT TRANSFER SIMULATIONS IN CUMNAS FILMS WITH DIFFERENT THICKNESSES**

The heat transfer simulations were performed using finite elements software Comsol Multiphysics. Due to the ratio of the ≈ 30 μm excitation spot diameter and the 10's of nm thicknesses of the CuMnAs films, we used the approximation of 1-dimensional heat diffusion from the film to the substrate. The pump laser was simulated as 150 fs long rectangular shape constant power density heat source in CuMnAs layer. GaP substrate absorption was neglected. The density was considered to be constant and ambient temperature was set to 15 K. The average temperature in CuMnAs layer was used as the readout parameter.

As a starting point, we compared results of simulations when spatially homogeneous and exponential decay absorption profiles were considered, respectively. For these two cases, we found no significant difference in the obtained results for a sample containing thin (8 nm) CuMnAs film. For the thicker (60 nm) film, there were some differences but they were not substantial in the explored time range of hundreds and thousands of ps, where the heat dissipation takes place (see Fig. 4c). Therefore, we used the homogeneous excitation conditions in all further simulations.

We simulated the experimental results shown in Fig. 4 taking into account the temperature-dependent heat capacity and heat conductivity – see Fig. 9. We used a Debye model for the specific heat capacity. For GaP we used values of Debye temperature $T_D$ = 445 K and heat capacity at 300 K of 430 J kg$^{-1}$ K$^{-1}$ [39]. The corresponding values for CuMnAs are not known, yet. Therefore, we used Debye temperature $T_D$ = 275 K and heat capacity at 300 K of 300 J kg$^{-1}$ K$^{-1}$ for a similar antiferromagnetic material CuMnSb [40]. The temperature dependencies of heat capacities (normalized to the values at 300 K) are shown in Fig. 9a. The heat conductivity of GaP at 300 K is 75 W m$^{-1}$ K$^{-1}$ [41,42]. For CuMnAs, we estimated the value based on conductivity in the out-of-plane direction $10^5$ S m$^{-1}$ and Wiedemann–Franz–Lorenz law [43] as 0.732 W m$^{-1}$ K$^{-1}$. We used the normalized temperature dependence of copper heat conductivity as an approximation of CuMnAs temperature dependence, as shown in Fig. 9b. In the simulations, the effective heating power densities, which are shown in the inset of Fig. 5a, were estimated from the incident laser pump pulse peak power density, the measured absorptances of individual films (see Fig. 6d) and the film thicknesses. The results of the simulations are shown in Fig. 5a. For a characterization of a speed of the heat dissipation dynamics, we used an effective thermal relaxation time $\tau$, which corresponds to time needed to lower the CuMnAs temperature increase $\Delta T$ to $1/e$ of the initial value.

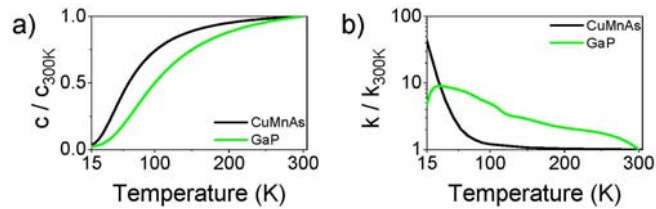

**Figure 9: Temperature dependence of heat capacity *C* (a) and heat conductivity *K* (b) for GaP and CuMnAs**; the displayed values are normalized to the values at 300 K. Note that in (b) logarithmic y-scale is used.